\RequirePackage{ifpdf}
\ifpdf % We are running pdfTeX in pdf mode
\documentclass[pdftex]{sigma}
\else
\documentclass{sigma}
\fi

\newcommand{\mbf}[1]{{\boldsymbol {#1} }}

\newcommand{\Leq}{\preceq}
\newcommand{\Le}{\prec}
\newcommand{\Geq}{\succeq}
\newcommand{\Ge}{\succ}

\newcommand{\vv}{v}
\newcommand{\ee}{{\rm e}}
\newcommand{\ii}{{\rm i}}

\newcommand{\vx}{\mathbf{x}}
\newcommand{\vn}{\mathbf{n}}

\newcommand{\cf}{\check f}
\newcommand{\cP}{\check P}

\newcommand{\cPT}{{\cal PT}}

\newcommand{\cH}{{\cal H}}
\newcommand{\cE}{{\cal E}}
\newcommand{\cV}{{\cal V}}

\newcommand{\cR}{{\cal R}}
\newcommand{\cS}{{\cal S}}
\newcommand{\cC}{{\cal C}}

\renewcommand{\Lambda}{C}
\newcommand{\eps}{\epsilon}

\newcommand{\vy}{\mathbf{y}}
\newcommand{\vs}{\mathbf{s}}

\newcommand{\vz}{\mathbf{z}}

\newcommand{\vm}{\mathbf{m}}

\newcommand{\ve}{\mathbf{e}}
\newcommand{\vE}{\mathbf{E}}
\newcommand{\vmu}{\mbf{\hat\mu}}

\newcommand{\vxi}{\mbf{\xi}}

\newcommand{\eq}{\begin{equation*}}
\newcommand{\eqend}{\end{equation*}}

\newcommand{\eeq}{\begin{equation}}
\newcommand{\eeqend}{\end{equation}}

\newcommand{\eqa}{\begin{eqnarray}}
\newcommand{\eqaend}{\end{eqnarray}}

\newcommand{\nonueqa}{\begin{eqnarray*}}
\newcommand{\nonueqaend}{\end{eqnarray*}}

\newcommand{\bma}[1]{\begin{array}{#1}}
\newcommand{\ema}{\end{array}}

\newcommand{\Ref}[1]{\eqref{#1}}

\newcommand{\R}{{\mathbb R}}
\newcommand{\C}{{\mathbb C}}
\newcommand{\Z}{{\mathbb Z}}

\begin{document}

\allowdisplaybreaks

\renewcommand{\PaperNumber}{031}

\FirstPageHeading

\renewcommand{\thefootnote}{$\star$}

\ShortArticleName{On Calogero--Sutherland Type Systems}

\ArticleName{Singular Eigenfunctions of Calogero--Sutherland\\
Type Systems and How to Transform Them\\ into Regular
Ones\footnote{This paper is a contribution to the Vadim Kuznetsov
Memorial Issue `Integrable Systems and Related Topics'. The full
collection is available at
\href{http://www.emis.de/journals/SIGMA/kuznetsov.html}{http://www.emis.de/journals/SIGMA/kuznetsov.html}}}

\Author{Edwin LANGMANN} \AuthorNameForHeading{E. Langmann}

\Address{Theoretical Physics, KTH Physics, AlbaNova, SE-106 91 Stockholm, Sweden}

\Email{\href{mailto:langmann@kth.se}{langmann@kth.se}}
\URLaddress{\url{http://www.theophys.kth.se/~langmann/}}

\ArticleDates{Received November 02, 2006, in f\/inal form January
29, 2007; Published online February 26, 2007}

\Abstract{There exists a large class of quantum many-body systems
of Calogero--Suther\-land type where all particles can have
dif\/ferent masses and coupling constants and which nevertheless
are such that one can construct a complete (in a certain sense)
set of exact eigenfunctions and corresponding eigenvalues,
explicitly. Of course there is a catch to this result: if one
insists on these eigenfunctions to be square integrable then the
correspon\-ding Hamiltonian is necessarily non-hermitean (and thus
provides an example of an exactly solvable $\cPT$-symmetric
quantum-many body system), and if one insists on the Hamiltonian
to be hermitean then the eigenfunctions are singular and thus not
acceptable as quantum mechanical eigenfunctions. The standard
Calogero--Sutherland Hamiltonian is special due to the existence
of an integral operator which allows to transform these singular
eigenfunctions into regular ones.}


\Classification{81U15; 33C50; 05E05}

\renewcommand{\thefootnote}{\arabic{footnote}}
\setcounter{footnote}{0}

\section{Prologue}
In summer 2003 Stefan Rauch informed me that a colleague of us
would soon visit him in Link\"oping and thus pass through
Stockholm, and he suggested that I should invite him for a
seminar: this colleague was also working on quantum integrable
systems of Calogero--Sutherland (CS) type \cite{Cal2,Su2} and I
would want to talk to him.  I thus got to know Vadim Kuznetsov on
September 1 in 2003, short before his seminar where he explained
to us their work on the $Q$-ope\-rator and separation of variables
for the CS model \cite{KMS}. Stefan Rauch was right: not only did
I f\/ind the seminar very inspiring and interesting, but also in
the discussion afterward I learned about a (for me) new motivation
for studying CS type systems which, since then, has inf\/luenced
and helped me signif\/icantly in my work on this topic. By
`learned' I mean here that some way of looking at things, known
only from hear-say before and very dif\/ferent from my own,
suddenly `sank in' and made me see things from a new angle. This
was a rare experience for me which, I felt, was due to a special
channel of communication and mutual understanding I had when
talking to Vadim. Vadim was enthusiastic about our science (not
only then but in all the, unfortunately too few, instances I had
the privilege to meet him), his enthusiasm spread to me, and this
was a main reason why I soon afterward returned to my work on the
CS system \cite{EL1,EL0} and thus stumbled over results much more
extensive and beautiful than what I had expected and for which I
am very grateful (these results were announced in \cite{EL4,EL5},
and the f\/inal version of the detailed paper on this, by which I
intend to replace \cite{EL3}, is still in the stage of being
`nearly f\/inished' since 20 months. I miss Vadim's interest and
kind encouragement, but I intend to f\/inish it soon, I want to
make it as good as I can, and when eventually f\/inished it will
be to a large degree Vadim's paper since it would not exist
without him).

%Vadim's scientific background was very different from mine, he had a
%way of unterstanding things which was very different from mine, but
%due to this I found it very inspiring and enjoyable to interact with
%him.

In the present paper I will present a story which I had carried
with me since quite a while (my old draft on it which I use when
writing this paper was dated in October 2000) without being able
to fully make sense of it. It is about a seemingly provocative
result, closely related to this above-mentioned method to
construct explicit formulas for eigenfunctions of such
systems~\cite{EL5}, and I discussed it with Vadim on March 2 (or
3) in 2005, shortly after a seminar to which Vadim had invited me
to explain them my approach. I brought up this story in response
to a question of Vadim, and after our discussion I understood its
place and it f\/inally made sense to me. I~remember Vadim liked
this story, and this is why I choose it for the present article.
Moreover, as will become clear later, this story is related in
spirit to B\"acklund transformations and the $Q$-operator which
have played a central role in Vadim's research; see e.g.\
\cite{KS,KPR}.

My plan for the rest of this paper is as follows. In the next
section I will summarize some known facts about CS systems which
are relevant to my story. In Section~\ref{sec31} I will present
the `strange' exact eigenfunctions for a large class of
generalized CS-type systems, and in Sections~\ref{sec32} and
\ref{sec33} I clarify their nature by pointing out a natural
interpretation in the context of $\cPT$-symmetric mechanics
\cite{PT}: they provide non-hermitean $\cPT$-invariant quantum
many-body Hamiltonians with real spectrum and computable
eigenvalues and eigenfunctions.  In Section~\ref{sec4} I will
explain that these strange eigenfunctions for the CS
dif\/ferential operator are interesting and useful since they can
be transformed into regular eigenfunctions. This provides an
alternative approach to results previously obtained in \cite{EL5}.
I will end with a few remarks in Section~\ref{sec5}. Technical
details and some more technical proofs are deferred to three
appendices.

\section{Background}
\label{sec2}

\subsection{CS model and Jack polynomials}
\label{sec21} The \textit{Calogero--Sutherland (CS) model}
\cite{Cal2,Su2} is def\/ined by the dif\/ferential operator
\begin{gather}
\label{CS} H_N = - \sum\limits_{j=1}^N\frac{\partial^2}{\partial
x_j^2} +  2\lambda (\lambda-1) \sum\limits_{j<k}
\frac{1}{4\sin^2\bigl(\frac12(x_j-x_j)\bigr)}
\end{gather}
with $x_j \in[-\pi,\pi]$ coordinates on a circle, $N=2,3,\ldots$,
and $\lambda>0$.  This dif\/ferential operator essentially
def\/ines a quantum mechanical model of $N$ identical particles
moving on a circle and interacting with the $1/\sin^2$ two-body
potential, and the parameter $\lambda$ determines the coupling
strength.  The CS model is famous among theoretical physicists
since it can be solved exactly: the dif\/ferential operator $H_N$
has exact eigenfunctions of the form
\begin{gather}
\label{Psin} \Psi_{\vn}(\vx) = \Psi_0(\vx)P_{\vn}(\vz)
\end{gather}
which are labeled by partitions $\vn$ of length $N$, i.e.,
$\vn=(n_1,n_2,\ldots,n_N)$ with $n_j$ integers such that
\begin{gather*}
n_1\geq n_2\geq \cdots \geq n_N\geq 0,
\end{gather*}
and where $P_\vn$ are certain symmetric polynomials of degree
$|\vn|=\sum\limits_{j=1}^N n_j$ in the variables
\begin{gather}
\label{zj} z_j = \ee^{\ii x_j}
\end{gather}
with
\begin{gather}
\label{Psi0} \Psi_0(\vx) =
\prod\limits_{j<k}\sin\biggl(\dfrac12(x_j-x_k)\biggr)^\lambda
\end{gather}
the groundstate eigenfunction. Moreover, the corresponding exact
eigenvalues have the following remarkably simple form,
\begin{gather}
\label{En} E_{\vn} = \sum\limits_{j=1}^N \biggl( n_j +
\dfrac12(N+1-2j)\lambda\biggr)^2.
\end{gather}
The functions $P_{\vn}(\vz)$ are called {\em Jack polynomials} and
can
  be characterized, somewhat informally, as follows
\begin{gather*}
P_{\vn}(\vz) = \Biggl(\sum\limits_{P}
z_1^{Pn_{1}}z_2^{Pn_{2}}\cdots z_{N}^{Pn_{N}}\Biggr) + \mbox{
lower order terms}
\end{gather*}
where the sum is over all distinct permutations $P$ of
$(n_1,n_2,\ldots,n_N)$; see e.g.\ \cite{St} for a precise
characterization.

The following two remarks are on technicalities which can be often
safely ignored, but I regard them useful as preparation for our
discussion of the generalized CS models in the next section.

\begin{remark}
For non-integer $\lambda$ our def\/inition of $\Psi_0(\vx)$ above
is only complete in the wedge $-\pi<x_1<x_2<\dots < x_N<\pi$ where
we require it to be real (since the phase factors $(-1)^\lambda$
obtained by permuting the arguments are obviously ambiguous).
There is one natural method to f\/ix this ambiguity by analytical
continuation: extend the def\/inition of $\Psi_0(\vx)$ to the
other regions by continuing the $x_j$ to the complex plane,
$x_j\to x_j+\ii \eps j$ with $\eps>0$, performing the necessary
permutation, and then taking the limit $\eps\to 0$.
\end{remark}

\begin{remark}
\label{rem1} It is worth mentioning another seemingly technical
point which, however, will play an important role for us later:
our `def\/inition' of the CS model above by $H_N$ in \Ref{CS} was
somewhat vague since, to be precise, this quantum mechanical model
is def\/ined by a particular self-adjoint operator on the Hilbert
space $L^2([-\pi,\pi]^N)$. Such an operator is only uniquely
def\/ined if one specif\/ies an operator domain, and the latter we
did implicitly by characterizing the eigenfunctions. To see that
this is not only a technicality we note that the CS model is
specif\/ied by the parameter $\lambda$, but the coupling constant
\begin{gather*}
\gamma= 2\lambda(\lambda-1)
\end{gather*}
in \Ref{CS} is invariant under $\lambda\to 1-\lambda$. Thus in the
coupling regime $-1/2 < \gamma<0$ one and the same CS
dif\/ferential operator corresponds to two dif\/ferent CS models.
In fact, the dif\/ferential operator in \Ref{CS} has many more
self-adjoint extensions. We will later encounter other examples
where one and the same dif\/ferential operator def\/ines
dif\/ferent Hilbert space operators.
\end{remark}

%\begin{remark}
%The Jack polynomials above are even defined for negative values of
%$\lambda$, and even though these cases are not interesting from a
%physics point of view (since the eigenfunctions are not square
%integrable they cannot describe quantum mechanical states which are
%physically meaningful) they still have mathematical interest.
%\end{remark}

We note that these Jack polynomials are just one example in a
whole zoo of symmetrical polynomials related to eigenfunctions of
CS-type models and which naturally generalize the classical
orthogonal polynomials (like Hermite, Laguerre, Jacobi, \ldots) to
the many-variable case. These polynomials were, and still are,
extensively studied in mathematics, and many beautiful results
have been discovered; see e.g.\ \cite{Dunkl} and references
therein. It was only due to discussions with Vadim that I became
aware that my results could be relevant in that context, and it
was he who convinced me to write \cite{EL5} where I could obtain
explicit formulas for the Jack polynomials which (to my knowledge)
were not know before. One aim of the present paper is to give an
alternative derivation and interpretation of these formulas. To be
specif\/ic I restrict my discussion to the Jack polynomials and
only mention in passing that all results allow for a (rather)
straightforward generalization to the other symmetric polynomials
mentioned above~\cite{HL1,HL2}.

\subsection{A generalized CS model}
\label{sec22} There exists a generalization of the CS model
describing distinguishable particles and which is partially
solvable in the sense that is has a groundstate eigenfunctions and
corresponding groundstate energy which can be computed explicitly.
This model is def\/ined by the dif\/ferential operator
\begin{gather}
\label{CS1} \cH_N = -
\sum\limits_{j=1}^N\frac1{M_j}\frac{\partial^2}{\partial x_j^2} \;
+ \sum\limits_{j<k} \frac{\gamma_{jk} }{ 4
\sin^2\biggl(\dfrac12(x_j-x_k)\biggr) }
\end{gather}
with arbitrary mass parameters $M_j/2>0$ and the coupling
constants
\begin{gather}
\label{gamjk} \gamma_{jk} = (M_j+M_k)\lambda(M_j M_k\lambda -1) ,
\end{gather}
and it has the following exact groundstate
\begin{gather}
\label{Phi0} \Phi_0(\vx) =
\prod\limits_{j<k}\sin\biggl(\dfrac12(x_k-x_j)\biggr)^{\lambda M_j
  M_k}
\end{gather}
with corresponding groundstate energy
\begin{gather}
\label{cE0} \cE_0 = \frac{\lambda^2}{12}\Biggl(\Biggl(
\sum\limits_{j=1}^N M_j\Biggr)^3 -\sum\limits_{j=1}^N M_j^3
\Biggr),
\end{gather}
i.e., $\cH_N\geq \cE_0$ and
\begin{gather}
\cH_N\Phi_0(\vx) = \cE_0\Phi_0(\vx) . \label{HPhi}
\end{gather}
This fact is known since quite a while (see e.g.\ \cite{Sen}), but
due to its importance for us a short proof is given in
Appendix~\ref{appA}. This fact plays a twofold role in our story:
f\/irstly, it suggests that the Hamiltonian in \Ref{CS1} might
have further exactly computable eigenstates, and the search for
these led to the result presented in the next section, and
secondly, it implies, as special case, the following result which
will play an important role for us in Section~\ref{sec4}:

\begin{lemma}
\label{lem1} The function
\begin{gather*}
F_N(\vx;\vy) = c \, \ee^{\ii
P\sum\limits_{j=1}^N(x_j-y_j)}\frac{\prod\limits_{1\leq
  j<k\leq N} \sin((1/2)(x_j-x_k))^\lambda \prod\limits_{1\leq j<k\leq N}
  \sin((1/2)(y_j-y_k))^\lambda}{ \prod\limits_{j,k=1}^N
  \sin((1/2)(x_j-y_k))^\lambda}
\end{gather*}
with $x_j$, $y_j$ complex variables, $c$ and $P$ arbitrary constants,
and the CS Hamiltonian $H_N=H_N(\vx)$ in \Ref{CS}, obey the
following identity,
\begin{gather}
H_N(\vx)F_N(\vx,\vy) = H_N(\vy)F_N(\vx,\vy) . \label{remId}
\end{gather}
\end{lemma}

(The proof is given in Appendix~\ref{appC0}.)

\medskip

As was pointed out to me by Vadim, the identity in \Ref{remId} is
also implied by a well-known generating function of the Jack
polynomials which is closely related to the functions $F_N$; see
Proposition~2.1 in \cite{St}.  We will use this identity to
construct an integral transform $\hat F_N$ in Section~\ref{sec4}
which commutes with the CS Hamiltonian and which, for this very
reason, will be useful for us. This integral transform is similar
to the $Q$-operator def\/ined and exploited in \cite{KMS}, and the
usefulness of def\/ining this integral transform $\hat F_N$ was
suggested to me by Vadim.

\section{Strange exact eigenfunctions of CS-type system}
\label{sec3} In this section we construct and discuss singular
eigenfunctions of the CS-type dif\/ferential ope\-ra\-tor in
\Ref{CS1}. We note that this result highlights the importance of
square integrability in quantum mechanics: if one ignores this
condition it is actually easy to construct explicit eigenfunctions
for a large variety of models. We f\/irst will discuss the
simplest class of eigenfunction (Section~\ref{sec31}).
Section~\ref{sec32} contains an interpretation of these as
eigenfunctions of a non-hermitean Hilbert space operator with
purely real spectrum. We also sketch a generalization of this
construction illustrating that one and the same dif\/ferential
operator can def\/ine a large class of dif\/ferent Hilbert space
operators.

\subsection{Construction of singular eigenfunction}
\label{sec31} We now construct exact eigenfunctions of the
dif\/ferential operator in \Ref{CS1} with arbitrary coupling
constants $\gamma_{jk}$. These eigenfunctions are labeled by
integer vectors $\vn\in\Z^N$ and are linear combinations of the
monomials
\begin{gather}
\label{cf} \cf_{\vn}(\vz) \, : = \, z_1^{n^+_1}z_2^{n^+_2}\cdots
  z_N^{n^+_N},\qquad n^+_j = n^{\phantom +}_j + s^{\phantom +}_j
\end{gather}
where $\vs\in \R^N$ is arbitrary for now. It is important to note
that these eigenfunctions are {\it not} symmetric, i.e., not
invariant under permutations of the particles. Moreover, we will
f\/ind that these eigenfunctions are only well-def\/ined in a
region contained in the following domain,
\begin{gather*}
\Omega_N = \bigl\{ \vz\in\C^N \bigm| |z_1|<|z_2|<\cdots
<|z_N|\bigr\}
\end{gather*}
and thus are, in particular, not elements in the Hilbert space
$L^2([-\pi,\pi]^N)$.\footnote{In Section~\ref{sec32} we will
discuss an alternative interpretation of these functions as
elements in this Hilbert space.}  To state our result we will also
need the following partial ordering of integer vectors
\begin{gather*}
%\label{part}
\vm\Leq \vn\; \Leftrightarrow\; m_j+m_{j+1}+\cdots +m_N \leq
n_j+n_{j+1}+\cdots +n_N \qquad \forall \; j=1,2,\ldots,N ,
\end{gather*}
and we will use the special integer vectors
\begin{gather*}
\vE_{jk}=\ve_j-\ve_k
\end{gather*}
with $\ve_j$ the standard basis vectors in $\Z^N$, i.e.,
$(\ve_j)_\ell = \delta_{j\ell}$ for all $j,\ell=1,2,\ldots,N$.  We
will also use the notation $\delta_{\vn}(\vm)$ for the Kronecker
delta in $\Z^N$, i.e.,
\begin{gather*}
\delta_{\vn}(\vm) = \delta_{n_1,m_1}\delta_{n_2,m_2}\cdots
\delta_{n_N,m_N} .
\end{gather*}
Moreover, we will use the following subset of $\Z^N$:
\begin{gather*}
\Lambda^N_-\, = \, \Biggl\{\vmu=
\mbox{$\displaystyle\sum\limits_{j<k}$}\mu_{jk}\vE_{jk}
\biggm|\mu_{jk}=0,1,2,\ldots \Biggr\} .
\end{gather*}
Note that $\vn+\vmu\Leq \vn$ for all $\vmu\in\Lambda^N_-$ and
$\vn\in\Z^N$.

We start our construction by the following simple observation:

{\samepage
\begin{lemma}
\label{lem2} In the region $\Omega_N$ the differential operator in
\Ref{CS1}, for arbitrary $\gamma_{jk}$, acts on the functions in
\Ref{cf} as follows,
\begin{gather}
\label{T} \cH_N \cf_{\vn} = E_{\vn} \cf_{\vn} -
\sum\limits_{j<k}\gamma_{jk} \sum\limits_{\nu=1}^\infty \nu
\cf_{\vn + \nu \vE_{jk}}
\end{gather}
where
\begin{gather}
\label{En1}
 E_{\vn} = \sum\limits_{j=1}^N \frac{(n_j+s_j)^2}{M_j}.
\end{gather}
\end{lemma}

}

\begin{proof}
Using
\begin{gather*}
-\ii \frac{\partial}{\partial x_j} = z_j \frac{\partial}{\partial
z_j} \qquad \mbox{and}\qquad \frac1{4\sin^2(1/2(x_j-x_k))} = -
\frac{{z_j}/{z_k}}{(1- {z_j}/{z_k})^2} = -
\sum\limits_{\nu=1}^\infty\nu \biggl( \frac{z_j}{z_k}\biggr)^\nu
\end{gather*}
for $|z_j/z_k|<1$, we can write the dif\/ferential operator in
\Ref{CS1} on the domain $\Omega_N$ as follows,
\begin{gather*}
\cH_N = \sum\limits_{j=1}^N \frac1{M_j} \left(
z_j\frac{\partial}{\partial z_j}\right)^2 - \sum\limits_{j<k}
\gamma_{jk} \sum\limits_{\nu=1}^\infty \nu \left( \frac{z_j}{z_k}
\right)^\nu.
\end{gather*}
This implies the result.
\end{proof}

Equation \Ref{T} shows that the action of $\cH_N$ on the functions
$\cf_{\vn}$ has triangular structure in the following sense:
$\cH_N \cf_{\vn}$ is a linear superposition of functions
$\cf_{\vm}$ with $\vm\Leq \vn$. This suggests that~$\cH_N$ should
have eigenfunctions $\cP_{\vn}$ which can be expanded as linear
combinations of functions~$\cf_{\vm}$, $\vm\Leq \vn$, and with
eigenvalues $E_{\vn}$. This turns out to be the case, and one
f\/inds that the corresponding expansion coef\/f\/icients can be
computed by diagonalizing a triangular matrix if a certain
non-degeneracy condition is fulf\/illed, and that this matrix is
simple enough to allow for a fully explicit construction of these
eigenfunctions. It is also possible to establish a necessary
condition for absolute convergence of this series. To summarize:

\begin{proposition}
\label{prop1} Let $E_{\vn}$ be as in \Ref{En1},
\begin{gather}
\label{bnm} b_{\vn}(\vm) = E_{\vm}-E_{\vn} = \sum\limits_{j=1}^N
\frac1{M_j} (m_j-n_j)(m_j+n_j + 2 s_j),
\end{gather}
and $\vn\in\Z^N$, $\vs\in\R^N$ and $M_j>0$ such that
\begin{gather}
\label{cond} \left| b_{\vn}(\vn+\vmu) \right|>\Delta \qquad
\forall\; \vmu\in\Lambda^N_-
\end{gather}
for some $\Delta>0$. Then the function
\begin{gather}
\label{cP} \cP_{\vn} = \cf_{\vn} + \sum\limits_{\vm\Le\vn}
\alpha_{\vn}(\vm) \cf_{\vm}
\end{gather}
with
\begin{eqnarray}
\label{alph} \alpha_{\vn}(\vm) = \delta_{\vn}(\vm) +
\sum\limits_{s=1}^\infty \prod\limits_{r=1}^s\left(
\sum\limits_{j_r<k_r}\gamma_{j_rk_r} \sum\limits_{\nu_r=1}^\infty
\nu_r\right)
\frac{\delta_{\vn}\Bigl(\vm-\sum\limits_{r=1}^s\nu_r\vE_{j_rk_r}\Bigr)}{\prod\limits_{r=1}^s
b_{\vn}\Bigl(\vn+\sum\limits_{\ell=1}^r\nu_\ell\vE_{j_\ell
k_\ell}\Bigr)}
\end{eqnarray}
is an exact eigenfunction of the differential operator in
\Ref{CS1} corresponding to the eigenvalue $E_{\vn}$ in \Ref{En1}.
Moreover, the infinite series defining this function is absolutely
convergent in the region where
\begin{gather}
\label{cond1} \sum\limits_{j<k} |\gamma_{jk}| \frac{|z_j/z_k|}{(1-
|z_j/z_k|)^2} < \Delta.
\end{gather}
\end{proposition}
\begin{proof}
Inserting the ansatz \Ref{cP} in the eigenvalue equation
$(\cH_N-E)\cP_{\vn}=0$, using \Ref{T}, and renaming one summation
variable we obtain
\begin{gather*}
\sum\limits_{\vm\Leq \vn} \Biggl( (E_{\vm}-E)\alpha_{\vn}(\vm)
-\sum\limits_{j<k}\gamma_{jk}\sum\limits_{\nu=1}^\infty \nu
\alpha_{\vn}(\vm-\nu\vE_{jk}) \Biggr)\cf_{\vm} = 0
\end{gather*}
with $\alpha_{\vn}(\vn)=1$. Since the functions $\cf_{\vm}$ are
linearly independent we conclude that this is true if and only if
\begin{gather*}
(E_{\vm}-E)\alpha_{\vn}(\vm)-\sum\limits_{j<k}\gamma_{jk}\sum\limits_{\nu=1}^\infty
\nu \alpha_{\vn}(\vm -\nu\vE_{jk}) = 0
\end{gather*}
for all $\vm\Leq\vn$. Since $\alpha_{\vn}(\vm)=0$ for $\vm>\vn$,
solving the latter equation amounts to computing a particular
eigenvector of a triangular matrix indexed by integer vectors, as
anticipated (see Appendix~\ref{appB}).  We conclude $E=E_{\vn}$
and that the coef\/f\/icients $\alpha_{\vn}(\vm)$ can be computed
from the following recursion relations
\begin{gather*}
\alpha_{\vn}(\vm) = \frac1{b_{\vn}(\vm)}
\sum\limits_{j<k}\gamma_{jk}\sum\limits_{\nu=1}^\infty \nu
\alpha_{\vn}(\vm -\nu\vE_{jk}),\qquad \alpha_{\vn}(\vn)=1
\end{gather*}
using the notation in \Ref{bnm}, provided that $b_{\vn}(\vm)\neq
0$ for all $\vm\Le\vn$. The latter is guaranteed by our assumption
in \Ref{cond}. We can solve this recursion relation by iteration
(see Appendix~\ref{appB}). We thus obtain
\begin{gather*}
\alpha^{\phantom s}_{\vn}(\vm) = \sum\limits_{s=0}^\infty
\alpha^{(s)}_{\vn}(\vm)
\end{gather*}
where
\begin{gather*}
\alpha^{(0)}_{\vn}(\vm)=\delta_{\vn}(\vm),\qquad
\alpha^{(s)}_{\vn}(\vm) = \frac1{b_{\vn}(\vm)}
\sum\limits_{j<k}\gamma_{jk}\sum\limits_{\nu=1}^\infty \nu
\alpha^{(s-1)}_{\vn}(\vm -\nu\vE_{jk})
\end{gather*}
for $s=1,2,\ldots$. The formula in \Ref{alph} is obtained by
computing
\begin{gather*}
\alpha^{(s)}_{\vn}(\vm) = \frac1{b_{\vn}(\vm)}
\sum\limits_{j_s<k_s}\gamma_{j_sk_s}\sum\limits_{\nu_s=1}^\infty
\nu_s \frac1{b_{\vn}(\vm-\nu_s \vE_{j_sk_s})}
\sum\limits_{j_{s-1}<k_{s-1}}\gamma_{j_{s-1}k_{s-1}}
\sum\limits_{\nu_{s-1}=1}^\infty \nu_{s-1}
\\[-.5ex] \phantom{\alpha^{(s)}_{\vn}(\vm) =}
{}\times\frac1{b_{\vn}\Bigl(\vm-\sum\limits_{\ell=s-1}^s \nu_\ell
\vE_{j_\ell k_\ell}\Bigr)}
\sum\limits_{j_{s-2}<k_{s-2}}\gamma_{j_{s-2}k_{s-2}}
\sum\limits_{\nu_{s-2}=1}^\infty \nu_{s-2}
\frac1{b_{\vn}\Bigl(\vm-\sum\limits_{\ell=s-2}^s \nu_\ell
\vE_{j_\ell k_\ell}\Bigr)}
\\[-.5ex] \phantom{\alpha^{(s)}_{\vn}(\vm) =}
{}\times \cdots \frac1{b_{\vn}\Bigl(\vm-\sum\limits_{\ell=2}^s
\nu_\ell \vE_{j_\ell k_\ell}\Bigr)}
\sum\limits_{j_1<k_1}\gamma_{j_1k_1}\sum\limits_{\nu_1=1}^\infty
\nu_1 \delta_{\vn}\Bigl(\vm-\mbox{$\sum\limits_{r=1}^s \nu_r
\vE_{j_rk_r}$}\Bigr)
\end{gather*}
and replacing $\vm$ in the arguments of the $b_{\vn}$ by
$\vn+\sum\limits_{r=1}^s \nu_r \vE_{j_rk_r} $.

To see where the eigenfunctions $P_{\vn}$ are well-def\/ined
insert \Ref{alph} in \Ref{cP} and obtain
\begin{gather}
\label{cP1} \cP_{\vn}(\vz) = \vz^{\vn^+} +
\sum\limits_{s=1}^\infty \prod\limits_{r=1}^s\Biggl(
\sum\limits_{j_r<k_r}\gamma_{j_rk_r} \sum\limits_{\nu_r=1}^\infty
\nu_r\Biggr) \frac{1}{\prod\limits_{r=1}^s
b_{\vn}\Bigl(\vn+\sum\limits_{\ell=1}^r\nu_\ell\vE_{j_\ell
k_\ell}\Bigr)}\vz^{\left(\vn +
\sum\limits_{r=1}^s\nu_r\vE_{j_rk_r}\right)^+}
\end{gather}
where use the notation $\cf_{\vm}(\vz)=\vz^{\vm^+}$. We can
estimate this using \Ref{cond}, {\samepage\begin{gather*}
|\cP_{\vn}(\vz)|\leq |\vz^{\vn^+}|+ \sum\limits_{s=1}^\infty
\prod\limits_{r=1}^s\left(
\sum\limits_{j_r<k_r}|\gamma_{j_rk_r}|\sum\limits_{\nu_r=1}^\infty
\nu_r\right) \frac1{\Delta^s} |\vz^{\vn^+}| \prod\limits_{r=1}^s
\left|\frac{z_{j_r}}{z_{k_r}}\right|^{\nu_r}
\\[-.5ex] \phantom{|\cP_{\vn}(\vz)|}
{}=\sum\limits_{s=0}^\infty
\left(\sum\limits_{j<k}|\gamma_{jk}|\sum\limits_{\nu=1}^\infty \nu
\left|\frac{z_j}{z_k}\right|^\nu \frac1{\Delta} \right)^s
|\vz^{\vn^+}|
\end{gather*}}%
which implies absolute convergence of the series def\/ining
$P_{\vn}$ provided that the condition in \Ref{cond1} holds true.
\end{proof}

\begin{remark}
As the proof above shows, the condition in \Ref{cond1} is
suf\/f\/icient but far from necessary. We believe that, if
$b_{\vn}(\vm)\neq 0$ for all $\vm\Leq\vn$, then $P_{\vn}(\vz)$ is
well-def\/ined in all of $\Omega_N$.
\end{remark}
\begin{remark}
We call the eigenfunctions in Proposition~\ref{prop1} {\it
singular} since they seem to diverge as $z_j\to z_k$, at least for
positive coupling values: from a heuristic argument we expect that
they behave like
\begin{gather*}
\sim |z_j-z_k|^{(1/2)-\sqrt{(1/4)+\tilde\gamma_{jk}}} ,\qquad
\tilde\gamma_{jk} = \frac{M_jM_k}{M_j+M_k} \gamma_{jk}
\end{gather*}
as $z_j\to z_k$. This argument suggests that these eigenfunctions
are elements in $L^2([-\pi,\pi]^N)$ if all $\tilde \gamma_{jk}$
are negative and larger than $-1/4$.  It would be interesting to
prove that this was indeed the case since then these `singular'
eigenfunctions would be actually acceptable as quantum mechanical
wave functions.
\end{remark}
\begin{remark}
The idea that the (standard) CS dif\/ferential operator has
eigenfunctions as in \Ref{cP} was probably f\/irst used by Heckman
and Opdam; see e.g.\ \cite{HO}.
\end{remark}
\begin{remark}
Other remarkable special case of the generalized CS dif\/ferential
operator in~\Ref{CS1} and~\Ref{gamjk} allowing for polynomial
eigenfunctions are for $M_j=1$ for $1\leq j\leq n<N$ and
$M_j=-1/\lambda$ for $n<j\leq N$, as discovered in \cite{VFC} for
the case $n=N-1$ and explored for general $n$ in \cite{SV,SV1}. In
fact, all our results on the CS model eigenfunctions described
here can be extended to these deformed cases \cite{HL2}.
\end{remark}

The following two subsections can be skipped by readers not
interested in $\cPT$-symmetric quantum mechanics.

\subsection{Exactly solvable  $\cPT$-symmetric quantum many-body systems}
\label{sec32} We now show that the eigenfunctions in
Proposition~\ref{prop1} can be interpreted as solutions of an
(partially) exactly solvable $\cPT$-symmetric quantum many-body
system; see e.g.\ \cite{PT}.  For that we observe that complex
variables \Ref{zj} with $|z_j|=R^j$, $R>0$, can be obtained by
shifting the particle coordinates to the complex plane as follows,
\begin{gather*}
x_j\to x_j+\ii j \eps
\end{gather*}
and doing this replacement in the dif\/ferential operator in
\Ref{CS1} we obtain
\begin{gather}
\label{CS2} \cH_N = -
\sum\limits_{j=1}^N\frac1{M_j}\frac{\partial^2}{\partial x_j^2} \;
+ \sum\limits_{j<k} \frac{\gamma_{jk} }{ 4
\sin^2\biggl(\dfrac12(x_j-x_k - \ii [k-j]\eps )\biggr) } .
\end{gather}
It is interesting to note that this def\/ines an operator on the
Hilbert space $L^2([-\pi,\pi]^N)$ which is {\em not} hermitean but
rather obeys
\begin{gather*}
\cH_N^* = \overline{\cH_N^{\phantom *}}
\end{gather*}
where the star means Hilbert space adjungation and the bar complex
conjugation. However, this dif\/ferential operator is invariant
under the following combined parity~(${\cal P}$) and time
reversal~(${\cal T}$) transformation \cite{PT}:
\begin{gather*}
\cPT:\qquad x_j\to -x_j, \qquad \ii \to -\ii
\end{gather*}
implying $p_j = -\ii\partial/\partial x_j\to p_j$.  Moreover,
Proposition~\ref{prop1} provides exact eigenfunctions of this
dif\/ferential operators, and the corresponding eigenvalues are
indeed real, as they should be \cite{PT}:

\begin{corollary}
\label{cor2} The functions in \Ref{cP1} with
\begin{gather}
\vz^{\vm^+} = \ee^{\sum\limits_{j=1}^N ( \ii x_j - j
\eps)(m_j+s_j)}
\end{gather}
and $b_{\vn}(\vm)$ in \Ref{bnm} are exact formal\footnote{The
qualifier `formal' here means that questions of convergence are
ignored.}  eigenfunctions of the $\cPT$-symmetric Hamiltonian
$\cH_N$ in~\Ref{CS2} corresponding to the eigenvalue $E_{\vn}$
in~\Ref{En1}. Moreover, this eigenfunction is a well-defined
function in $L^2([-\pi,\pi]^N)$ provided that \Ref{cond1} holds
true for some $\Delta>0$ and
\begin{gather}
\label{cond3} \sum\limits_{j<k} |\gamma_{jk}|
\frac{\ee^{-2(k-j)\eps}}{(1+\ee^{-2(k-j)\eps})^2} <\Delta.
\end{gather}
\end{corollary}

\begin{proof}[Sketch of proof.] The only non-trivial part is to verify
  the suf\/f\/icient condition for square integrability: compute
\begin{gather*}
\int\limits_{[-\pi,\pi]^N}d^N x |\cP_{\vn}(\vx)|^2 =
\sum\limits_{\vm}|\alpha_{\vn}(\vm)|^2 \ee^{-2 (m_j+s_j)j\eps},
\end{gather*}
$\alpha_{\vn}(\vm)$ in \Ref{alph}, and majorize by a geometric
series, similarly as in the proof of Proposition~\ref{prop1}
above.
\end{proof}

It is interesting to note that for the CS dif\/ferential operator
in \Ref{CS} and $s_j$ in \Ref{sj} below, the condition in
\Ref{cond3} holds true automatically for all partitions $\vn$ and
suf\/f\/iciently large $\eps$ (e.g.\ any value $\eps\geq \log(R)$
with $R$ in \Ref{R1} will do), as shown in the next section (see
Lemma~\ref{lem3} and its proof in Appendix~\ref{appC1}).

If for such a $\cPT$-symmetric Hamiltonian the exact eigenfunction
in Corollary~\ref{cor2} exists for {\em all} integer vectors
$\vn$, then these eigenfunctions provide a complete basis in the
Hilbert space (it is not dif\/f\/icult to prove this, using that
the functions $\cf_{\vn}(\vz)$, $\vn\in\Z^N$, are a complete
basis). In this case we have an exactly solvable model, otherwise
we only might have an partially solvable model. However, we
believe that the condition in \Ref{cond1} can be relaxed and that
it is possible to compute all eigenfunctions explicitly, i.e., the
$\cPT$-symmetric Hamiltonian in \Ref{CS1} with eigenfunctions of
the form \Ref{cP} is always an exactly solvable model.

Note that we have obtained for the dif\/ferential operator in
\Ref{CS2} a family of eigenfunctions depending on parameters $s_j$
which we are free to choose. This is similar to what is well-known
for hermitean operators: a dif\/ferential operator does not
uniquely determine a self-adjoint operator, but there is usually a
whole family of self-adjoint extensions labeled by continuous
parameters; see e.g.\ \cite{RS2}. It is interesting to note that
one can easily construct a much larger family of $\cPT$-symmetric
Hamiltonians corresponding to the dif\/ferential operator in
\Ref{CS2}. In the following section we outline this construction,
but our discussion is somewhat sketchy. We only include it here in
the hope that it might be of interest in the context of
$\cPT$-symmetric quantum mechanics~\cite{PT}.

\subsection{Generalized $\cPT$-symmetric quantum many-body system}
\label{sec33} This section contains a sketch how to construct
eigenfunctions the the generalized CS dif\/ferential operator in
\Ref{CS1} depending on $N(N+1)/2$ parameters $s_j$ and
$\lambda_{jk}$, $1\leq j<k\leq N$ (the eigenfunctions in
Proposition \Ref{prop1} correspond to the special case
$\lambda_{jk}=0$).

Def\/ine
\begin{gather}
\label{Psi1} \Psi_0(\vx) =
\prod\limits_{j<k}\sin\biggl(\dfrac12(x_j-x_k-\ii[k-j]\eps)\biggr)^{\lambda_{jk}}
\end{gather}
with arbitrary real parameters $\lambda_{jk}$. Then the ansatz
\begin{gather*}
\Psi(\vx) = \Psi_0(\vx)\cP(\vz)
\end{gather*}
in the eigenvalue equation $\cH_N\Psi =E\Psi$ is equivalent to
$\cH'_N \cP = E'\cP$ with
\begin{gather*}
\cH'_N = -
\sum\limits_{j=1}^N\frac1{M_j}\frac{\partial^2}{\partial x_j^2} \;
+ \sum\limits_{j<k}\biggl( \frac{\gamma'_{jk} }{ 4
\sin^2((1/2)(x_j-x_k - \ii [k-j]\eps )) }\nonumber
\\ \phantom{\cH'_N =}
{}-\lambda_{jk}\cot\biggl(\dfrac12(x_j-x_k - \ii [k-j]\eps
)\biggr)\left(\frac1{M_j}\frac{\partial}{\partial x_j} -
\frac1{M_k}\frac{\partial}{\partial x_k} \right) \biggr),
\\
\gamma_{jk}'=\gamma_{jk}
-\frac{M_j+M_k}{M_jM_k}\lambda_{jk}(\lambda_{jk}-1)
\end{gather*}
and $E'=E-E_0$ with
\begin{gather}
\label{E0} E_0 =
\sum\limits_{j<k<\ell}\frac{1}{2M_j}\lambda_{jk}\lambda_{j\ell}
  -\sum\limits_{j\neq k}\frac1{4M_j}\lambda_{jk}^2.
\end{gather}
Note that for the special case where $\lambda_{jk}=\lambda
M_jM_k$, $\gamma_{jk}$ in~\Ref{gamjk}, $\eps=0$, the function
$\Psi_0(\vx)$ in~\Ref{Psi1} becomes identical to the one
in~\Ref{Phi0}, $\gamma_{jk}'=0$, and $E_0$ is identical to $\cE_0$
in \Ref{cE0}. We thus we recover~\Ref{HPhi}.  Another interesting
case is $M_j=1$, $\lambda_{jk}=\lambda$,
$\gamma_{jk}=2\lambda(\lambda-1)$ in which case $\cH_N'$ becomes
equal to the reduced Hamiltonian used by Sutherland in his
solution of the CS model~\cite{Su2}. By expanding the $\cot$- and
$1/\sin^2$-terms in power series one can compute the action of
$\cH'_N$ on the functions in \Ref{cf} which, again, is triangular,
and one then can construct eigenfunctions of the form \Ref{cP},
similar as in Section~\ref{sec31}.  One thus f\/inds the following
generalization of Proposition~\Ref{prop1}:

\begin{proposition}
\label{prop2} The differential operator in \Ref{CS1} has formal
eigenfunctions labeled by $\vn\in\Z^N$ and corresponding to the
eigenvalues
\begin{gather*}
E_{\vn} = \sum\limits_{j=1}^N\frac{(n_j+s_j)^2}{M_j} +
\sum\limits_{j<k}
\lambda_{jk}\biggl(\frac{(n_j+s_j)}{M_j}-\frac{(n_k+s_k)}{M_k}
\biggr)+ E_0
\end{gather*}
with $E_0$ in \Ref{E0}. These formal eigenfunctions are given by
\begin{gather*}
\Psi_{\vn}(\vx) = \Psi_0(\vx) \cP_{\vn}(\vz)
\end{gather*}
with $\Psi_0(\vx)$ in \Ref{Psi1} and $\cP_{\vn}(\vz)$ in \Ref{cP}
where the coefficients are determined by the following recursion
relations,
\begin{gather}
\label{alph1} (E_{\vn}-E_{\vm})\alpha_{\vn}(\vm) =
\sum\limits_{j<k}\sum\limits_{\nu=1}^\infty
  \biggl( \lambda_{jk}\biggl(\frac{(m_j+s_j+\nu)}{M_j} -\frac{(m_k+s_k
  -\nu)}{M_k}\biggr) -\nu\gamma_{jk}' \biggr) \alpha_{\vn}(\vm)
\end{gather}
with $\alpha_{\vn}(\vn)=1$.
\end{proposition}
Note that the recursion relations in \Ref{alph1} still have
triangular form and thus can, in principle, be solved by the
method explained in Appendix~\ref{appB} provided there are no
degeneracies, i.e.\ $E_{\vn}\neq E_{\vm}$ for all pertinent
$\vm\Leq \vn$. However, the resulting formula will be much more
involved and probably not very illuminating.

It is interesting to note that, if
\begin{gather*}
\frac{n_j+s_j}{M_j} > \frac{n_k+s_k}{M_k}\qquad \forall \; j<k ,
\end{gather*}
then the eigenvalues in \Ref{En} can be written in the following
simple form,
\begin{gather*}
E_{\vn} = \sum\limits_{j=1}^N \frac{p_j^2}{M_j},\qquad p_j = n_j +
\frac12 \sum\limits_{k<j}\lambda_{jk} - \frac12
\sum\limits_{k>j}\lambda_{jk} .
\end{gather*}

\section{A method to solve the CS model}
\label{sec4} In our discussion of the singular eigenfunctions all
generalized CS dif\/ferential operators in \Ref{CS1} could be
treated on equal footing. It thus seems natural to ask: is there
anything special about the CS dif\/ferential operator in \Ref{CS}
with regard to these singular eigenfunctions? To answer this
question we compare the eigenvalues in \Ref{En} and \Ref{En1} and
observe that, if we choose
\begin{gather}
\label{sj} s_j = \dfrac12 (N+1-2j)\lambda
\end{gather}
and restrict the integer vectors $\vn$ to partitions, then the
singular eigenfunctions of $H_N$ in \Ref{cP} not only are labeled
by the same quantum numbers but also have the very same
eigenvalues as its regular eigenfunctions in \Ref{Psin}. Moreover,
these singular eigenfunctions are well-def\/ined, i.e., the
condition in \Ref{cond} is automatically fulf\/illed. We now show
that this is no coincidence: \textit{The CS differential operator
$H_N$ is special since there exists an integral operator $\hat
F_N$ which commutes with it and which transforms its singular
eigenfunctions in \Ref{cP} as specified above into regular ones as
in \Ref{Psin}.} Thus we can obtain all the regular eigenfunctions
of the CS model by f\/irst constructing singular eigenfunctions
(which can be done explicitly), and then transforming them using
the operator $\hat F_N$. As shown below, in this way we recover
the explicit formulas for Jack polynomials derived previously in
\cite{EL5}.

In the following we make these statements more precise.  For that
we f\/irst def\/ine the class of functions $\cR_N$ and $\cS_N$ to
which the regular- and singular eigenfunctions of the CS
dif\/ferential operator belongs.

\begin{definition}
\label{def1} The regular domain $\cR_N$ of the CS dif\/ferential
operator in \Ref{CS} is the vector space of all functions of the
form $\Psi_0(\vx)P(\vz)$ with $\Psi_0(\vx)$ in \Ref{Psi0} and
$P(\vz)$ a symmetric function in the variables $z_j=\ee^{\ii x_j}$
which is analytic and bounded on the domain $|z_j|=1$ $\forall j$.
\end{definition}

\begin{definition}
\label{def2} The singular domain $\cS_{N}$ of the CS
dif\/ferential operator in \Ref{CS} is the vector space of all
Laurent series of the form
\begin{gather}
\label{cS0} \sum\limits_{\vmu\in\Lambda^N_-} a_{\vmu}
\cf_{\vn+\vmu}(\vz)
\end{gather}
for some f\/ixed $\vn\in\Z^N$, with $\cf_{\vm}$ in \Ref{cf}, $s_j$
in \Ref{sj}, and coef\/f\/icients $a_{\vmu}$ such that
\begin{gather}
\label{cS} \sum\limits_{\vmu\in\Lambda^N_-} |a_{\vmu}|
\prod\limits_{j<k}^N R^{-\mu_{jk}(k-j)} <\infty
\end{gather}
for some $R>1$ which is suf\/f\/iciently large.
\end{definition}

\begin{remark}
The convergence condition in \Ref{cS} is chosen such that the
series in \Ref{cS0} is absolutely convergent in the region where
$|z_j| =R^j$ $\forall j$, and it then obviously is convergent for
$|z_j|\geq R^j$ $\forall j$.  We introduce this parameter $R>1$ to
avoid technicalities: our crude arguments in Appendix~\ref{appC1}
show that the following value of $R$ is suf\/f\/iciently large,
\begin{gather}
\label{R1} R =  \max(2,N(N-1)|\lambda-1|/8),
\end{gather}
but we suspect that any value $R>1$ would do.
\end{remark}

We now summarize the properties of the singular eigenfunctions of
the CS dif\/ferential operator following from
Proposition~\ref{prop1}.  As mentioned, the non-trivial part is
that the conditions in \Ref{cond1} are automatically fulf\/illed.

\begin{lemma}
\label{lem3} For all partitions $\vn$ of length $N$ and $s_j$ as
in \Ref{sj} the series $\cP_{\vn}$ in \Ref{cP} and \Ref{alph} for
$\gamma_{jk}=2\lambda(\lambda-1)$ and $M_j=1$ (independent of
$j,k$) are well-defined singular eigenfunctions of the CS
differential operator \Ref{CS} corresponding to the eigenvalue
\Ref{En}:
\begin{gather*}
\cP_{\vn}\in\cS_{N} \, \mbox{ and }\,  H_N\cP_{\vn}=E_{\vn}
\cP_{\vn} .
\end{gather*}
\end{lemma}

(The proof is given in Appendix~\ref{appC1}.)

\medskip

We now explain how to construct the operator $\hat F_N$.  The idea
is to use the function $F_N(\vx,\vy)$ in \Ref{remId} as integral
kernel and def\/ine
\begin{gather}
\label{hF1} \hat F_N(\cP)(\vx) = \int\limits_{\cC}\frac{d^N
y}{(2\pi)^N} F(\vx,\vy)\cP(\vxi),\qquad \xi_j = \ee^{\ii y_j}
\end{gather}
for suitable choices of the constants $P$ and $c$, and the
integration domain $\cC$. Then Lemma~\ref{lem1} guarantees that
this operator commutes with $H_N$, and it maps eigenfunctions of
$H_N$ into eigenfunctions of $H_N$ without changing the
eigenvalues, even though the character of these eigenfunctions is
very dif\/ferent.  A straightforward computation (deferred to
Appendix~\ref{appC2}) then shows how $c$, $P$, and $\cC$ need to
be chosen to get a simple and meaningful result, and by inspection
one f\/inds the domain and range of this operator. To be precise:

\begin{lemma}
\label{lem4} The prescription
\begin{gather}
\label{hF} \hat F_N (\cP)(\vx) =\Psi_0(\vx) \prod\limits_{j=1}^N
\left(\oint_{|\xi_j|= R^j}\frac{d\xi_j}{2\pi\ii\xi_j}\right)
\frac{\prod\limits_{1\leq j<k\leq
N}(1-\xi_j/\xi_k)^\lambda}{\prod\limits_{j,k=1}^N(1-z_j/\xi_k)^\lambda}
\prod\limits_{j=1}^N \xi_j^{-s_j}\cP(\vxi)
\end{gather}
with $s_j$ in \Ref{sj} and $R>1$ sufficiently large, defines an
operator commuting with the CS differen\-tial operator \Ref{CS}
and mapping its singular- to its regular domain:
\begin{gather*}
\hat F_N(H_N \cP)(\vx) = H_N \hat F_N(\cP)(\vx)
\end{gather*}
and
\begin{gather*}
\hat F_N(\cP)\in\cR_N \, \mbox{ for all }\, \cP\in \cS_N.
\end{gather*}
\end{lemma}

(The proof is given in Appendix~\ref{appC2}.)

\medskip

Combining Lemmas \ref{lem3} and \ref{lem4} above we can conclude
that the functions $\hat F_N(\hat P_{\vn})(\vx)$ are regular
eigenfunctions of the CS dif\/ferential operator $H_N$ to the
eigenvalues in \Ref{En}, and they are of the form
\begin{gather*}
\hat F_N(\hat P_{\vn})(\vx) = \Psi_0(\vx)P_{\vn}(\vz).
\end{gather*}
One can prove that $P_{\vn}(\vz)$ is actually a symmetric
polynomial of degree $\sum\limits_{j=1}^N n_j$ and that
$P_{\vn}(\vz)$ is equal to the corresponding Jack polynomial, up
to normalization; see \cite{EL0} and \cite{HL2} for details.
Interchanging summation and the integral transform we thus obtain
the following representation of the Jack polynomials as a linear
superposition of the special symmetric polynomials $f_{\vn}=\hat
F_N(\cf_{\vn})$:

\begin{proposition}
\label{prop3} The Jack polynomial labeled by the partition $\vn$
is proportional to
\begin{gather*}
%\label{Jack}
P_{\vn}(\vz) = \sum\limits_{\vm\Leq \vn} \alpha_{\vn}(\vm)
f_{\vn}(\vz)
\end{gather*}
where
\begin{gather*}
%\label{fn}
f_{\vm}(\vz) = \prod\limits_{j=1}^N \left( \oint_{|\xi_j|=R^j}
\frac{d\xi_j}{2\pi\ii\xi_j} \xi_j^{m_j}\right)
\frac{\prod\limits_{1\leq j<k\leq
N}(1-\xi_j/\xi_k)^\lambda}{\prod\limits_{j,k=1}^N(1-z_j/\xi_k)^\lambda},
\\
\alpha_{\vn}(\vm) = \delta_{\vn}(\vm) + \sum\limits_{s=1}^\infty
[2\lambda(\lambda-1)]^s \prod\limits_{r=1}^s\Biggl(
\sum\limits_{j_r<k_r} \sum\limits_{\nu_r=1}^\infty \nu_r\Biggr)
\frac{\delta_{\vn}\Bigl(\vm-\sum\limits_{r=1}^s\nu_r\vE_{j_rk_r}\Bigr)}{\prod\limits_{r=1}^s
b_{\vn}\Bigl(\vn+\sum\limits_{\ell=1}^r\nu_\ell\vE_{j_\ell
k_\ell}\Bigr)}
\end{gather*}
and $b_{\vn}(\vm)=E_{\vn}-E_{\vm}$ with $E_{\vn}$ in \Ref{En}.
\end{proposition}

\begin{remark}
A non-trivial point in the result above is possible degeneracies:
if the partition $\vn$ is such that there is no other partition
$\vm$ such that $\sum\limits_j m_j = \sum\limits_j n_j$ and
$E_{\vn}=E_{\vm}$ then the corresponding eigenfunction of the CS
model is (essentially) unique and the symmetric polynomials thus
only can be equal to the corresponding Jack polynomial. However,
if this is not the case an additional argument is needed.
\end{remark}

\begin{remark} It is interesting to note that, in the special case $\lambda=1$, the
result above gives the following integral representation of the
Schur polynomials,
\begin{gather*}
S_{\vn}(\vz) = \prod\limits_{j=1}^N \Biggl(\,
\oint\limits_{|\xi_j|=R^j} \frac{d\xi_j}{2\pi\ii\xi_j}
\xi_j^{n_j}\Biggr) \frac{\prod\limits_{1\leq j<k\leq
N}(1-\xi_j/\xi_k)}{\prod\limits_{j,k=1}^N(1-z_j/\xi_k)};
\end{gather*}
this is a simple consequence of results in \cite{McD} (see
\cite{HL2}); see also Appendix~B in \cite{AFMO}.
\end{remark}

\begin{remark}
Integral representations of the Jack polynomials involving
integral operators somewhat similar to ours were previously
obtained in \cite{MY,AMOS}.
\end{remark}

\section{Epilogue}
\label{sec5} Yuri Suris at a meeting in Rome in May 2001 pointed
out to me that the (elliptic generalization) of the identity in
Lemma~\ref{lem1} looks very much like the B\"acklund
transformation for the classical Calogero--Sutherland system, but
I did not absorb this comment at that time. In January 2006 in
Rome again it eventually `sank in' and I eventually learned about
Stefan Rauch's pioneering work on this \cite{W}. I thus understand
now that it was not a coincidence that it was through Stefan that
I got to know Vadim who has played a central role in developing
the idea of quantum B\"acklund transformations (see
e.g.~\cite{KS}) which, in the special case of the CS model, is
based on the very same identity. I stumbled over this identity by
coincidence when working on a project where I studied a quantum
f\/ield theory model (initially with Alan Carey) with the aim to
learn more about the so-called fractional quantum Hall ef\/fect
\cite{EL1}, and through the positive feedback and encouragement I
received from the integrable system's community I started to
explore its consequences, and now I f\/ind myself working more and
more on problems related to the mathematical theory of special
functions. This happened to a large degree due to Vadim. He has
encouraged me and given me a few glimpses on some of his visions
for which I am very grateful and which will probably continue to
have a strong inf\/luence on my work. There will be probably more
papers in the future where Vadim's name should be mentioned~\ldots

\appendix

\section{Groundstate of the generalized CS model}
\label{appA} As discussed in the text, the Schr\"odinger operator
in \Ref{CS1} is a partially solvable system in the sense that its
exact groundstate and corresponding groundstate energy can be
found explicitly. In this appendix we give a precise formulation
and proof of this fact.

\begin{lemma}\label{lemma1}
Let $\Phi_0(\vx)$ be as in \Ref{Phi0} and
\begin{gather*}
Q^{\pm}_j = \pm \frac{\partial}{\partial x_j} + \cV_j \qquad
\mbox{with}\qquad \cV_j = \frac{\partial\log\Phi_0(\vx)}{\partial
x_j}
\end{gather*}
for arbitrary complex parameters $\lambda$ and $M_j$. Then
\begin{gather*}
\sum\limits_{j=1}^N \frac1{M_j} Q_j^+ Q^-_j = \cH_N - \cE_0
\end{gather*}
with $\cH_N$ the differential operator in \Ref{CS1} and $\cE_0$
the constant in \Ref{cE0}, and this implies \Ref{HPhi}.  In
particular, if all $M_j$ are positive then $\cH_N$ defines a
self-adjoint operator on the Hilbert space $L^2([-\pi,\pi]^N)$
bounded from below by $\cE_0$ (Friedrichs extension; see e.g.\
{\rm \cite{RS2}}), and $\Phi_0$ is its groundstate.
\end{lemma}

\begin{proof} We compute
\begin{gather*}
\cV_j = \sum\limits_{k\neq j} M_j M_k\lambda\phi(x_j-x_k) \mbox{
with } \phi(z) : = \dfrac12\cot\biggl(\dfrac{z}2\biggr)
\end{gather*}
and thus, using
\begin{gather*}
\phi'(z) = -\frac14 -\phi(z)^2
=-\frac14\sin^{-2}\biggl(\dfrac{z}2\biggr),
\end{gather*}
we obtain by straightforward computations
\begin{gather*}
\sum\limits_{j=1}^N \frac1{M_j} Q_j^+ Q^-_j = \sum\limits_{j=1}^N
\frac1{M_j} \left(-\frac{\partial^2}{\partial x_j^2} + \cV_j^2 +
\frac{\partial\cV_j}{\partial x_j} \right) = \cH_N - (*)
\end{gather*}
with $\cH_N$ in \Ref{CS1} and the reminder terms
\begin{gather*}
(*) = \lambda^2 \Biggl( -\sum\limits_{j}\sum\limits_{k\neq
j}\sum\limits_{\ell\neq j,k} M_jM_kM_\ell
\phi(x_j-x_k)\phi(x_j-x_\ell) + \frac14
\sum\limits_j\sum\limits_{k\neq j} M_j M_k^2 \Biggr).
\end{gather*}
The non-trivial part of the result thus is that the three-body
terms in $(*)$ add up to a constant. To see this we symmetrize the
f\/irst sum in $(*)$ above as follows,
\begin{gather*}
-2 \sum\limits_{j<k<\ell} M_jM_kM_{\ell}
[\phi(x_j-x_k)\phi(x_j-x_\ell) +  \phi(x_k-x_j)\phi(x_k-x_\ell) +
  \phi(x_\ell-x_j)\phi(x_\ell-x_k)].
\end{gather*}
This allows us to use the following trigonometric identity
\begin{gather*}
\phi(x_j-x_k)\phi(x_j-x_\ell) +  \phi(x_k-x_j)\phi(x_k-x_\ell) +
\phi(x_\ell-x_j)\phi(x_\ell-x_k) = -\frac14
\end{gather*}
and obtain
\begin{gather*}
(*) = \frac{\lambda^2}4\Biggl( 2\sum\limits_{j<k<\ell}
M_jM_kM_\ell + \sum\limits_{j<k} M_j M_k(M_j+M_k) \Biggr),
\end{gather*}
and by straightforward computations we f\/ind $(*)=\cE_0$ as in
\Ref{cE0}.

Since obviously $Q_j^-\Phi_0=0$ for all $j$ this proves
\Ref{HPhi}. Moreover, $Q_j^+$ is the hermitean conjugate of
$Q_j^-$, and if all $M_j>0$ then $\sum\limits_j (1/M_j)Q_j^+Q_j^-$
therefore def\/ines obviously a non-negative, self-adjoint
operator. This implies $\cH_N\geq \cE_0$, and thus $\Phi_0$ is
groundstate of $\cH_N$.
\end{proof}

\section{Proofs}
\label{appC}
\subsection{Proof of Lemma~\ref{lem1}}
\label{appC0} The identity in \Ref{HPhi} holds obviously true for
arbitrary real parameters $M_j$ and complex variables $x_j$ (see
Appendix~\ref{appA}).

We double the degrees of freedom and replace $N$ by $2N$. Setting
$ M_j=1$, $M_{N+j}=-1$, and $x_{N+j}=y_j$ for $j=1,2,\ldots,N$ we
f\/ind that $\cH_{2N}=H_N(\vx)-H_N(\vy)$ (note that all
interaction terms mixing the variables $x_j$ and $y_j$ are zero
since $(M_j+M_{N+k})=0$ for $j,k=1,2,\ldots,N$; we also slightly
abuse notation here), $\Phi_0(\vx,\vy)=F_N(\vx,\vy)$ for $P=0$,
and $\cE_0=0$. Thus the identity in \Ref{HPhi} implies \Ref{remId}
for $P=0$ and $c=1$.

We now observe that, due to translation invariance, $H_N(\vx)$ has
the following simple dependence on the center of mass coordinate
$X=\sum\limits_{j=1}^N x_j/N$,
\begin{gather*}
H_N(\vx) = -\frac{\partial^2}{\partial X^2} + \cdots
\end{gather*}
where the dots are terms independent of $X$ (they only depend on
the relative coordinates $x_j-x_N$ for $N=1,2,\ldots,N-1$). This
implies that the identity in \Ref{remId} is invariant under the
transformation
\begin{gather*}
F_N(\vx,\vy) \to c \, \ee^{\ii P' (X-Y)}F_N(\vx,\vy)
\end{gather*}
for arbitrary constants $P'=PN$ and $c$ (the invariance under
multiplication with $c$ is trivial, of course).

\subsection{Proof of Lemma~\ref{lem3}}
\label{appC1} We observe that
\begin{gather*}
b_{\vn}(\vn + \mbox{$\sum\limits_{j<k}\mu_{jk}\vE_{jk}$}) =
\sum\limits_{j=1}^N \Biggl(\, \sum\limits_{k=j+1}^N
\mu_{jk}[n_j-n_k+(k-j)\lambda] +
\Biggl[\,\sum\limits_{k<j}\mu_{kj} -
\sum\limits_{k>j}\mu_{jk}\Biggr]^2\Biggr)
\end{gather*}
which shows that \Ref{cond} holds true with $\Delta = 2\lambda$.
By a simple computation we f\/ind that \Ref{cond1} is equivalent
to
\begin{gather*}
\sum\limits_{j<k}\frac{|z_j/z_k|}{\left(1- |z_j/z_k|\right)^2} <
\frac{1}{|\lambda-1|}.
\end{gather*}
Replacing $|z_j|$ by $R^j$ we f\/ind that \Ref{cS} is equivalent
to
\begin{gather}
\label{RR} \sum\limits_{j<k}\frac{R^{j-k}}{\left(1- R^{j-k}
\right)^2} < \frac{1}{|\lambda-1|},
\end{gather}
and this holds true for suf\/f\/iciently large values of $R>1$,
e.g.\ by the following rough estimates
\begin{gather*}
\mbox{l.h.s. of \Ref{RR} } \leq \sum\limits_{j<k}\frac{R^{-1}}4 =
\frac{R^{-1} N(N-1)}8,
\end{gather*}
assuming $R\geq 2$, we f\/ind that the $R$ given in \Ref{R1} will
do.

\subsection{Proof of Lemma~\ref{lem4}}
\label{appC2} We express the function $F_(\vx,\vy)$ in terms of
the variables $z_j=\ee^{\ii x_j}$ and $\xi_j=\ee^{\ii y_j}$ (using
$ \sin(x/2)= z^{1/2}(1-1/z)/(2\ii)$), and by a straightforward
computation we f\/ind
\begin{gather*}
 F_N(\vx,\vy) = c\, \Psi_0(\vx)(2\ii)^{\lambda(N^2-N(N-1)/2)}
 \Biggl(\prod\limits_{j=1}^N \Bigl( z_j^{P+N\lambda/2} \xi_j^{-P-
 (N+1-2j)\lambda/2 -N\lambda/2}\Bigr) \Biggr)
 \\ \phantom{F_N(\vx,\vy) =}
 {}\times \frac{\prod\limits_{1\leq j<k\leq
 N}(1-\xi_j/\xi_k)^\lambda}{\prod\limits_{j,k=1}^N(1-z_j/\xi_k)^\lambda}
 \prod\limits_{j=1}^N \xi_j^{-s_j}.
\end{gather*}
To get rid of an awkward constant and the non-analytical (for
non-integer $\lambda$) factors $z_j$ we choose
\begin{gather*}
c=(2\ii)^{\lambda (N(N-1)/2-N^2)}\qquad \mbox{and}\qquad
P=-\dfrac{\lambda  N}2.
 \end{gather*}
 This yields
\begin{gather*}
 F_N(\vx,\vy) =  \Psi_0(\vx)  \frac{\prod\limits_{1\leq j<k\leq
N}(1-\xi_j/\xi_k)^\lambda}{\prod\limits_{j,k=1}^N(1-z_j/\xi_k)^\lambda}
\prod\limits_{j=1}^N \xi_j^{-\lambda(N+1-2j)/2}.
\end{gather*}
From this we see that if we choose the integration paths as
\begin{gather*}
\cC: \quad y_j = \phi_j -\ii j \eps,\qquad -\pi\leq \phi_j\leq
\pi\quad \mbox{for}\quad j=1,2,\ldots,N
\end{gather*}
then $\xi_j = R^j \ee^{\ii \phi_j}$ with $R=\ee^{\eps}>0$,
 and thus we f\/ind that \Ref{hF1} is
identical with what is given in~\Ref{hF}.

It is important to note that $\hat F_N(\cf_{\vn})$ for monomials
in \Ref{cf} is well-def\/ined if and only if $s_j$ is chosen as in
\Ref{sj} (modulo integers, of course), and $\cS_N$ is the natural
domain for $\hat F_N$ since it contains all linear superpositions
of these $\cf_{\vn}(\vz)$ which are absolutely convergent on $\cC$
(recall~\Ref{cS0}, $|\vz^{\vm}|=\prod\limits_{j}R^{j m_j}$ on
$\cC$, and \Ref{cS}): writing $\hat F_N(\cP)(\vx) =
\Psi_0(\vx)P(\vz)$,
\begin{gather*}
P(\vz) = \prod\limits_{j=1}^N \Biggl(\,\oint\limits_{|\xi_j|=
R^j}\frac{d\xi_j}{2\pi\ii\xi_j}\Biggr) \frac{\prod\limits_{1\leq
j<k\leq
N}(1-\xi_j/\xi_k)^\lambda}{\prod\limits_{j,k=1}^N(1-z_j/\xi_k)^\lambda}
\prod\limits_{j=1}^N \xi_j^{-s_j}\cP(\vxi)
\end{gather*}
we can estimate on $|z_j|=1$,
\begin{gather*}
|P(\vz)|\leq \frac{\prod\limits_{1\leq j<k \leq
    N}(1-R^{j-k})^\lambda}{\prod\limits_{k=1}^N(1- R^{-k})^{N\lambda}}
    \prod\limits_{j=1}^N R^{-j s_j}|\cP(\vxi)|
\end{gather*}
and thus prove that $\hat F_N(\cP)(\vx)$ is in the regular domain
$\cR_N$ provided $\cP$ is in the singular domain~$\cS_N$.

\section{Explicit diagonalization of triangular matrices}
\label{appB} In this Appendix we explain the method to explicitly
diagonalize a non-degenerate triangular matrix which we use in the
main text.  The matrix in the main text is indexed by integer
vectors, but to simplify our notation here we label the matrix
elements by integers $J,K\in\Z$, i.e., we write the matrix as
$A=(A_{JK})_{J,K\in\Z}$.

We are interested in f\/inding the eigenvector
$\vv=(\vv_J)_{J\in\Z}$ of this matrix $A$ corresponding to the
eigenvalue $E=A_{LL}$. Our assumptions about the matrix are as
follows,
\begin{gather*}
%\label{A}
A_{JK}\neq 0\quad\mbox{only if}\quad  K\Geq J,\qquad
A_{JJ}-A_{LL}\neq 0 \qquad \forall J \Ge L .
\end{gather*}
The symbol `$\Geq$' here can mean any partial ordering of the
integers.  We write $a_K = A_{KK}$ for the eigenvalues of this
matrix.

The $\vv_J$ are determined by the eigenvalue equation
\begin{gather*}
\sum\limits_{K\Geq J} A_{JK}\vv_K = E \vv_J
\end{gather*}
for $J\Leq L$, where $\vv_K=0$ for $K\Ge L$ and $\vv_L=1$. For
$J=L$ we get $E=A_{LL}=a_L$, and for $J\Le L$
\begin{gather*}
\vv_J = \frac1{(a_L-a_J)}\sum\limits_{K\Ge J} A_{JK}\vv_K
\end{gather*} which gives well-def\/ined recursion relations due to our
non-degeneracy assumption. We can solve this by iteration:
\begin{gather*}
\vv^{\phantom s}_J = \sum\limits_{s=0}^\infty \vv_J^{(s)}
\end{gather*}
where
\begin{gather*}
\vv_J^{(0)} = \delta^{\phantom s}_{JL},\qquad \vv_J^{(s)} =
\frac1{(a_L-a_J)}\sum\limits_{K\Ge J} A^{\phantom
s}_{JK}\vv^{(s-1)}_K\qquad \forall\; s>1
\end{gather*}
which is well-def\/ined since $\vv^{(s-1)}_K=0$ for $K\Ge L$.
This yields
\begin{gather*}
\vv_J  = \delta_{JL} + \sum\limits_{s=1}^\infty
\frac1{(a_L-a_J)}\sum\limits_{K_1\Ge J} A_{JK_1}
\frac1{(a_L-a_{K_1})}\sum\limits_{K_2\Ge K_1} A_{K_1K_2}\nonumber
\cdots
\\ \phantom{\vv_J  =}{}\times
\frac1{(a_L-a_{K_{s-1}})}\sum\limits_{K_s\Ge K_{s-1}}
A_{K_{s-1}K_s}\delta_{K_s,L} .
\end{gather*}
We thus get the following fully explicit formula for the
components of the eigenvector,
\begin{gather*}
\vv_J = \delta_{JL} + \sum\limits_{s=1}^\infty \sum\limits_{K_s\Ge
K_{s-1}\Ge \cdots
  \Ge
  K_1\Ge J} \frac{A_{JK_1}A_{K_1K_s}\cdots A_{K_{s-1}K_s}
  \delta_{K_sL}}{(a_L-a_{K_1})(a_L-a_{K_2})\cdots (a_L-a_{K_s})} .
\end{gather*}
Note that we keep the sum inf\/inite only for simplicity of
notation, but it actually has only a~f\/inite number of non-zero
terms: convergence is no problem here, of course.

\subsection*{Acknowledgements}
I would like to thank Orlando Ragnisco for organizing and inviting me
to several inspiring meetings in Rome and for showing me Ref.\ \cite{W}.
I thank Martin Halln\"as for useful comments on the manuscript and
Alexander Veselov for reading the paper and several helpful
comments. This work was supported by the Swedish Science Research
Council (VR) and the European Union through the FP6 Marie Curie
RTN {\em ENIGMA} (Contract number MRTN-CT-2004-5652).

\pdfbookmark[1]{References}{ref}
\LastPageEnding

\end{document}